# 3D MOKE spin-orbit torque magnetometer

Halise Celik[1], Harsha Kannan[1], Tao Wang[1], Alex R. Mellnik[2*], Xin Fan[3], Xinran Zhou[4],
Daniel C. Ralph[2], Matthew F. Doty[4], Virginia O. Lorenz[5] and John Q. Xiao[1†]

[1]*Department of Physics and Astronomy, University of Delaware, Newark, Delaware 19716, USA*

[2]*Department of Physics and Kavli Institute at Cornell for Nanoscale Science, Cornell University, Ithaca, New York 14853, USA*

[3] *Department of Physics and Astronomy, University of Denver, Denver, CO 80210, USA*

[4]*Department of Materials Science and Engineering, University of Delaware, Newark, Delaware 19716, USA*

[5] *Department of Physics, University of Illinois at Urbana-Champaign, Urbana, Illinois 61801, USA*

*Current Affiliation: Intel Corporation, Hillsboro, Oregon 97124, USA*

[†]jqx@udel.edu

**Abstract**

We demonstrate simultaneous detection of current driven antidamping-like and field-like spin-orbit torques in heavy metal/ferromagnetic metal bilayers by measuring all three magnetization components $m_x, m_y$, and $m_z$ using the vector magneto-optic Kerr effect. We have also implemented a self-calibration method to accurately determine the effective fields of spin-orbit torques. With this technique, we investigate the magnitude and direction of spin-orbit torques in a series of platinum/permalloy samples. The values found are in excellent agreement with results obtained via quadratic magneto-optic Kerr effect, planar Hall effect, and spin transfer ferromagnetic resonance measurements.

Spin–orbit coupling (SOC) driven phenomena such as the spin Hall effect (SHE) [1] and Rashba effect [2] enable manipulation of magnetization via electric current. Such electrical control of magnetization is crucial for future memory and logic spintronic devices in which memory and logic operations will be based on the magnetization direction [3–5]. By using electrical current to control the magnetization of nanoscale elements, it is possible to efficiently integrate magnetic functionalities into electronic circuits [6] and accelerate the technological development of high-performance and high-density magnetic storage devices [7–14].

In heavy metal (HM)/ferromagnetic metal (FM) bilayers, an electrical current will generate damping-like spin-orbit torque (SOT) and field-like spin-orbit torque (SOF), which will change the magnetization direction. In order to quantify the magnitude and the direction of magnetization reorientation due to SOT and SOF generated by the SHE and Rashba effects, electrical measurement techniques such as anomalous Hall effect and second-harmonic Hall effect measurements have been implemented for samples with perpendicular magnetic anisotropy [15]. The planar Hall effect (PHE) has also been used for samples with in-plane anisotropy [16]. However, these methods are second-order measurements and are susceptible to thermal effects or rectification effects due to other nonlinear processes that are common in magnetic materials. The torque values extracted from the measurements are also sensitive to fitting parameters like the anisotropy field [17]. Finally, it is not possible to measure both SOT and SOF terms simultaneously for the case of planar Hall effect measurements [15,16,18]. Spin-transfer-torque ferromagnetic resonance (STT FMR) [12] has also been used for samples with either perpendicular or in-plane anisotropy, but it suffers from possible intermixing between the spin pumping and the inverse spin Hall effect signals (ISHE) for very thin films [19].

Previously we have shown that normal incidence light can measure both current-induced out-of-plane magnetization reorientation by polar magneto-optic Kerr effect (MOKE) measurements [20] and in-plane magnetization reorientation by second-order (quadratic) MOKE measurements [21]. Such MOKE techniques do not suffer from electrical artifacts. More importantly, this approach allows self-calibration to extract the spin-orbit torque very accurately without any fitting parameters [17,20,21]. In ferromagnetic thin films, there have been several studies to determine the magnetization components vectorially [22,23]. For example, Ding *et al.* proposed a method to distinguish the pure longitudinal and polar Kerr contributions via two separate measurements interchanging the positions of a light source and a detector [24]. Yang *et al.* showed the detection of three magnetization components by changing the different relative orientations of the optical devices: polarizer, modulator, and analyzer [25]. However, these methods are complicated to implement and cumbersome in practice. As a simpler alternative, Keathley *et al.* used a scanning Kerr microscope equipped with a compact optical quadrant bridge polarimeter to measure in-plane vector hysteresis loops [26].

Here we present a vector MOKE spin-orbit torque magnetometer (vMTM) based on an optical quadrant bridge detector for first-order detection of current-induced SOT and SOF in HM/FM bilayers over a wide range of thicknesses. With this vMTM technique, in which light is at normal incidence, one can separate the linear and quadratic parts of the magnetization and determine all three components of the magnetization vector. We thus can measure both SOT and SOF components. We apply this method to measure SOT and SOF for a series of Platinum (Pt)/Permalloy ($Ni_{81}Fe_{19}$=Py) samples. We compare our results with measurements made using quadratic MOKE, the planar Hall effect, and spin transfer ferromagnetic resonance and find excellent agreement.

The Landau–Lifshitz–Gilbert–Slonczewski equation is usually used to describe the SOT and SOF generated from a current through the HM/FM bilayer [27]

$$\frac{d\vec{M}}{dt} = -\gamma\,\vec{M}\times\vec{H} + \frac{\alpha}{M_s}\vec{M}\times\frac{d\vec{M}}{dt} + a\vec{M}\times\vec{\sigma} + b\vec{M}\times(\vec{\sigma}\times\vec{M}), \qquad (1)$$

where σ is a unit vector for spin direction that is in-plane and orthogonal to the electric current, a and b describe the SOF and SOT, respectively. The effective fields corresponding to the SOF and SOT can be defined as $\vec{H}_{SOF} = -a\vec{\sigma}/\gamma$ and $\vec{H}_{SOT} = -b\vec{\sigma}\times\vec{M}/\gamma$, respectively.

The magneto-optic Kerr effect is the rotation of the polarization plane and change of polarization state of an electromagnetic wave as it reflects from a magnetized medium. As a result of MOKE the axis of polarization of linearly polarized light rotates (Kerr rotation angle $\theta_K$) and a slight ellipticity is introduced (Kerr ellipticity $\varepsilon_K$). These two quantities form the complex Kerr angle $\phi_K = \theta_K + i\varepsilon_K$. The Kerr rotation and ellipticity give a measure of the magnetization of the sample.

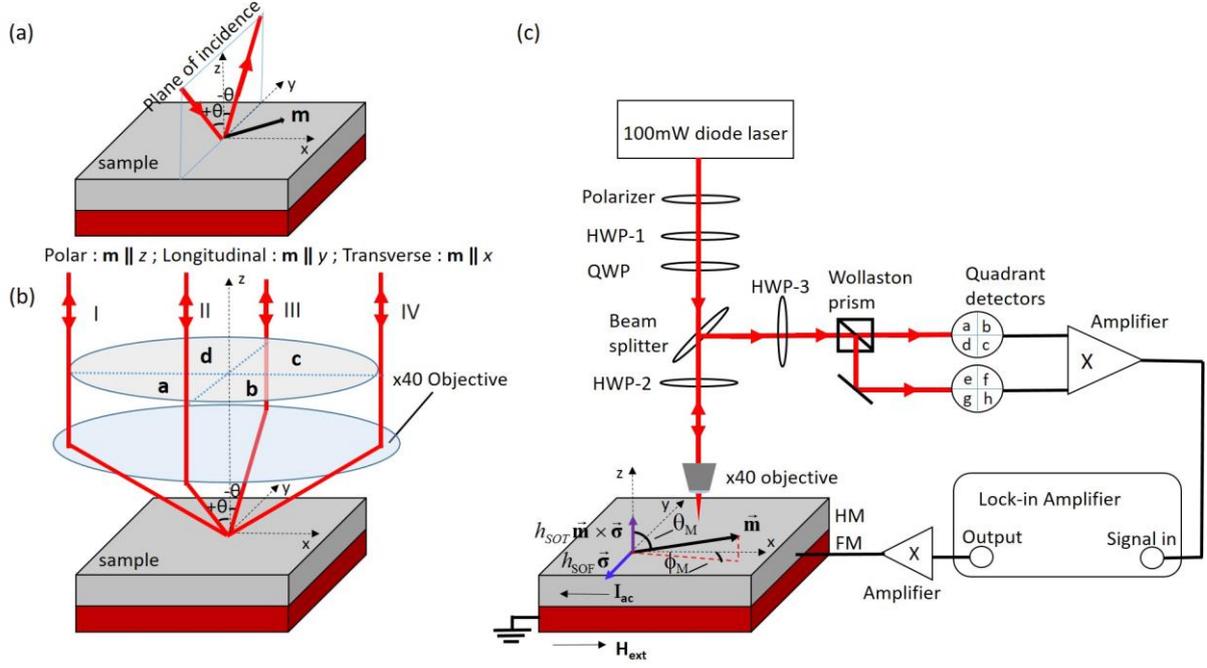

Figure 1. (a) Polar, longitudinal, and transverse MOKE geometries for a sample with magnetization $\boldsymbol{m}$. (b) Geometry of the optical quadrant bridge detection system. A 40X objective focusses light transmitted to and collimates light reflected from the sample. The reflected light is detected in spatial quadrants $a$, $b$, $c$, and $d$. By adding and subtracting signals from appropriate quadrants, one can isolate the in-plane MOKE response from the out-of-plane MOKE response. (c) Experimental setup for the optical detection of spin-orbit torques. HWP: half-wave plate, QWP: quarter-wave plate.

The magneto-optical properties of a material can be described by the permittivity tensor, $\varepsilon_{ij}$, which can be expanded in the components of the magnetization $\boldsymbol{m}$ acting on the material [28]:

$$\varepsilon_{ij} = \varepsilon_{ij}^{(0)} + K_{ijk}m_k + G_{ijkl}m_k m_l + \cdots, \qquad (2)$$

where the Einstein summation convention over the $x$, $y$, and $z$ coordinates is used. The dielectric tensor $\varepsilon_{ij}^{(0)}$ represents the components of the permittivity tensor in the absence of any magnetization $\boldsymbol{m}$, $K_{ijk}$ is the linear magneto-optic tensor, and $G_{ijkl}$ is the quadratic magneto-optic tensor, which corresponds to a second-order MOKE response, often referred to as quadratic MOKE [29]. The linear response can be separated into terms corresponding to relative orientations of the unit vector of the magnetization $\widehat{m}$, plane of incidence (POI) and sample plane (SP), with polar geometry corresponding to $\widehat{m} \parallel POI$ and $\widehat{m} \perp SP$, longitudinal to $\widehat{m} \parallel POI$ and $\widehat{m} \parallel SP$ and transverse to $\widehat{m} \perp POI$ and $\widehat{m} \parallel SP$. Longitudinal and polar MOKE alter the polarization of the incident light from plane to elliptically polarized with the major axis rotated (Kerr rotation) [30]. Transverse MOKE does not result in a change of the polarization of the incident light. It involves a change in reflectivity for $p$-polarized light [31].

Since the Kerr effect exists for any arbitrary direction of the magnetization, for oblique incidence the detected MOKE signal $\Psi(m)$ from a sample with magnetization $\boldsymbol{m}$ can be written as

$$\Psi(m) = \alpha_{polar}m_z + \gamma_{longitudinal}m_y + \delta_{longitudinal}m_x + \beta_{quadratic}m_x m_y \ldots, \qquad (3)$$

where the *z* direction is perpendicular to the magnetic film plane (see Figure 1a), the *y* direction is parallel to the plane of the incident polarization, and $\alpha_{polar}$, $\gamma_{longitudinal}$, $\delta_{longitudinal}$ and $\beta_{quadratic}$ are the coefficients for the polar, longitudinal, and quadratic MOKE responses, respectively.

It has been shown that polar and longitudinal signals can be separated by measuring the Kerr signal in two reversed geometries [32], as the polar signal does not change sign if the angle of incidence is reversed from $+\theta$ to $-\theta$ but the longitudinal signal does change sign. That is, polar MOKE is an even function and longitudinal MOKE is an odd function of the incident angle. Quadratic MOKE is also an even function of the incident angle as shown in ref [33]. Thus, light incident at $-\theta$ (e.g. travelling from ray III to ray II in Fig. 1b) has the same sign for polar and quadratic signals but opposite sign for longitudinal signals as light incident at $+\theta$ (e.g. travelling from ray II to ray III in Fig. 1b).

Using the even and odd dependence on the incident angle we are able to separate polar, longitudinal, and quadratic MOKE responses. Using a microscope objective with high numerical aperture (NA = 0.65 in our setup), we focus light across a wide range of incident angles from perpendicular to the sample plane to oblique [26]. The light reflected from the sample is measured in four quadrants as shown in Figure 1b. The reflection signal contains contributions from the polar, longitudinal, and quadratic responses, with the longitudinal contribution antisymmetric with incident angle. Thus, the response for angles of incidence $\theta$ with inward ($+\theta$) and outward ($-\theta$) propagation can be represented as

$$\varepsilon_K^{\pm\theta} = \varepsilon_K^P \pm \varepsilon_K^L + \varepsilon_K^Q, \qquad (4)$$

where $\varepsilon_K^{\pm\theta}$ are the Kerr ellipticities for the respective angles of incidence, and $\varepsilon_K^P, \varepsilon_K^L$, and $\varepsilon_K^Q$ are the ellipticities for the polar, longitudinal, and quadratic magneto-optic Kerr effects, respectively. By taking the sum of both inward ($\varepsilon_K^{+\theta}$) and outward ($\varepsilon_K^{-\theta}$) signals one obtains twice the sum of the polar and quadratic Kerr ellipticities, and by taking the difference one obtains twice the longitudinal Kerr ellipticity. In this way one in-plane magnetization component, $m_y$, can be measured. The other in-plane magnetization component, $m_x$, can be gathered by subtracting the right and the left halves of the beam. The Kerr rotation from the SOF term and other in-plane longitudinal component can be measured in this way. Measurement at 45 degree polarization can be performed to cancel the quadratic contribution, which enables us to determine the polar contribution, and in turn the SOT term. Details of the methodology for separating polar and quadratic responses can be found in our previous work [21].

A diagram of our vector MOKE setup is shown in Figure 1c. Collimated light from a 100mW diode laser at 785nm center wavelength goes through a Glan Taylor polarizer with an extinction coefficient of $\sim 10^{-4}$ to set the polarization. Two wave plates (HWP-1 and QWP-1) are used to compensate the birefringence of optical elements in the system to ensure the light has linear polarization at the sample. The angle of polarization is controlled with a half-wave plate (HWP-2) before being focused by a microscope objective of numerical aperture 0.65 on the sample. The reflected beam passes back through the objective and HWP-2 and is reflected by a 90/10 beam splitter. It goes through another half-wave plate (HWP-3) and the vertical and horizontal polarization components are split by a Wollaston prism. The intensity of the two components are balanced by adjusting HWP-3. The polarization components are detected by two quadrant photodiode detectors whose outputs are the sums or differences of various halves of the beams. The outputs of the detectors are subtracted from each other to achieve common mode rejection and doubling of the signal and then amplified. The signal is measured by a lock-in amplifier locked to the frequency of the ac current driving the sample. We have used the same set of samples we used in Ref [21], namely in-plane magnetized substrate/Pt(6 nm)/Py ($d_{Py}$) bilayers, with $d_{Py}$ ranging from 2 to 10 nm.

We apply an in-plane ac current, $I_{ac} \cos \omega t$, at 1733 Hz with $I_{ac} = 20$ mA along the x-axis to the sample. An external magnetic field $H_{ext}$ is applied along the *x-axis* to align the magnetization. The current-induced SOF and SOT rotate the magnetization within the sample plane (changing $\phi_M$) and perpendicular to the plane (changing $\theta_M$), respectively. The magnetization change due to current-induced torques for in-plane magnetized samples can be written in terms of two orthogonal effective magnetic field components $h_{SOF}$ and $h_{SOT}$,

$$\begin{cases} \Delta \phi_M = \frac{h_{SOF}}{H_{ext}+H_{a\parallel}} \\ \Delta \theta_M = \frac{h_{SOT}}{H_{ext}+H_{a\parallel}+M_s-H_{a\perp}} \end{cases}, \qquad (5)$$

where $H_{a\parallel}$ is the in-plane anisotropy field, $H_{a\perp}$ is the out of-plane anisotropy field, and $M_s$ is the saturation magnetization. For an ordinary transition-metal ferromagnet like Permalloy, the in-plane anisotropy is negligible and $M_s$ is much larger than any of the fields discussed here. Thus, for current-induced magnetization reorientation, the change in the polar MOKE signal (proportional to $\Delta \theta_M$) should be approximately independent of the applied field for $H_{ext} \ll M_s$, while the current-induced change in the longitudinal MOKE signal (proportional to $\Delta \phi_M$) should scale approximately as $1/H_{ext}$. Examples of experimental results from 50 μm x 50 μm Py(8)/Pt(6), where the numbers in parentheses are thicknesses in nanometers, with a 20mA bias current and 1mW laser power are shown in Figure 2.

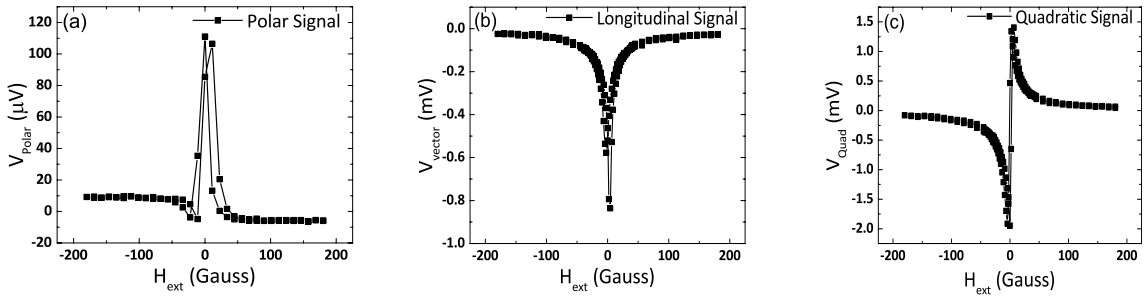

Figure 2. The current-induced (a) polar, (b) longitudinal and (c) quadratic magneto-optic Kerr response as a function of swept magnetic field $H_{ext}$ in Py(8)/Pt(6) bilayers.

Figure 2a shows the raw data for the Py(8)/Pt(6) polar term $(m_z)$ obtained using light at 45 degree polarization and taking the sum of all quadrants. It switches sign as the magnetization switches and is independent of $H_{ext}$ away from zero field. Figure 2b shows the longitudinal term $(m_y)$ at 0 degree polarization, which exhibits a $1/H_{ext}$ dependence. Figure 2c shows the raw data for a quadratic measurement $(m_x m_y)$. Since it is $m_x m_y$, it has $1/H_{ext}$ dependence but the symmetry is different than for the longitudinal measurement.

The magnitude of the SOT is determined through a self-calibration method explained in supplementary material. Using a simple parallel circuit model to account for the different resistivities of Pt and Py, we estimate that approximately 42% of the current flows through the Pt, yielding a current density in Pt of $j_{Pt} = 2.8 \times 10^{10} \, A/m^2$. By fitting the lines scans using quadrant detectors for the SOT signal and out-of-plane Oersted field, as seen in supplementary figure, we extract the SOT coefficient $\beta_T = \frac{h_{SOT}}{j_{Pt}} = 6.970 \pm 0.050 \, nm$. Assuming that the SOT arises from the spin Hall effect and using the equation $\theta_{SH} = \beta_T \left(\frac{2e}{\hbar}\right) \mu_0 M_s d_{Py}$, we determine a spin Hall angle of $\theta_\parallel = 0.086 \pm 0.007$ from vector MOKE for Pt, which is the same as that obtained with balanced detection, $\theta_\parallel = 0.086 \pm 0.004$ [21]. Here the parameters used are $\mu_0 M_s d_{Py} = 4.080 \, T.nm$.

To determine the magnitude of the SOF we perform a calibration by passing an ac current (500 mA) through a metallic wire (1 mm wide and 1 cm long) behind the sample that drives in-plane magnetization reorientation. This ac current generates an Oersted field of 70.700 ± 2.940 A/m. The distance from the sample to the wire is about 1.050 ± 0.050 mm. The magnitude of the SOF is extracted using a linear regression algorithm by comparing the SOF signal curve and the calibration curve shown in Figure 3a for Py(8)/Pt(6). In this example fitting, the ratio between the current-induced effective field and the calibration field is 2.490 ± 0.070, which corresponds to a current-induced field of 176.080 ± 5.310 A/m. After removing the 83.800 A/m Oersted field generated by the current in the sample, we obtain $h_{SOF}$ = 92.280 ± 5.310 A/m, which gives a spin Hall angle of $\theta_\perp = 0.054 \pm 0.003$. We also measure the other in-plane magnetization component, $m_x$, which is negligibly small as expected.

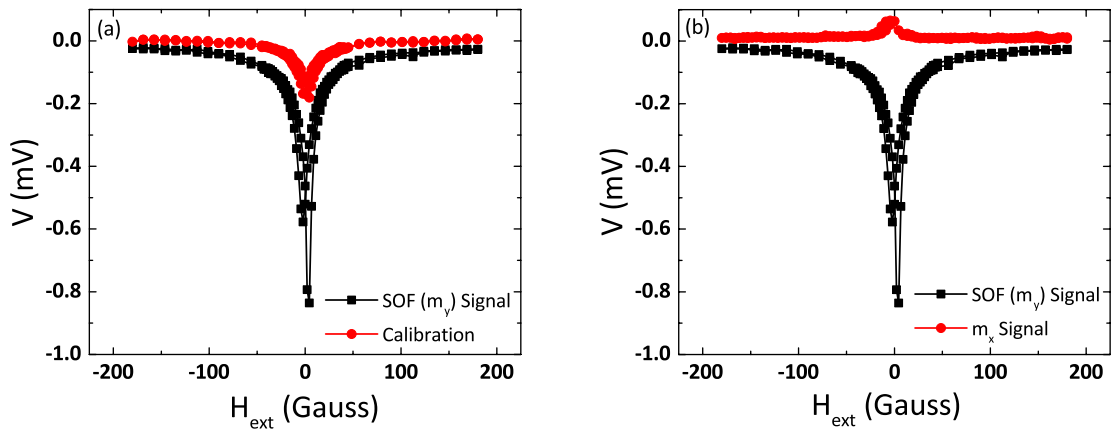

Figure 3. (a) Measured voltage from the lock-in amplifier as a function of the external magnetic field for Py(8)/Pt(6) when passing an ac current (20mA) through the sample (black squares) and an ac current (500mA) through a wire underneath the sample (red circles). (b) Comparison of signals in two longitudinal configurations, $m_y$ (black squares) and $m_x$ (red circles).

To further verify the accuracy of this method, we have extracted the spin Hall angle from SOF measurements for permalloy thicknesses $d_{Py}$ from 2-10 nm and compared the results with quadratic MOKE, spin-torque ferromagnetic resonance (ST-FMR), and planar Hall effect

(PHE) measurements. As seen in Figure 4a the quadratic and vector MOKE methods have excellent agreement. A comparison of the vector MOKE result with ST-FMR and PHE measurements is shown in Figure 4b. The four measurement techniques are in quantitative agreement. The ST-FMR measurement is performed as explained in ref [21] and the PHE measurement performed as in Ref. [16].

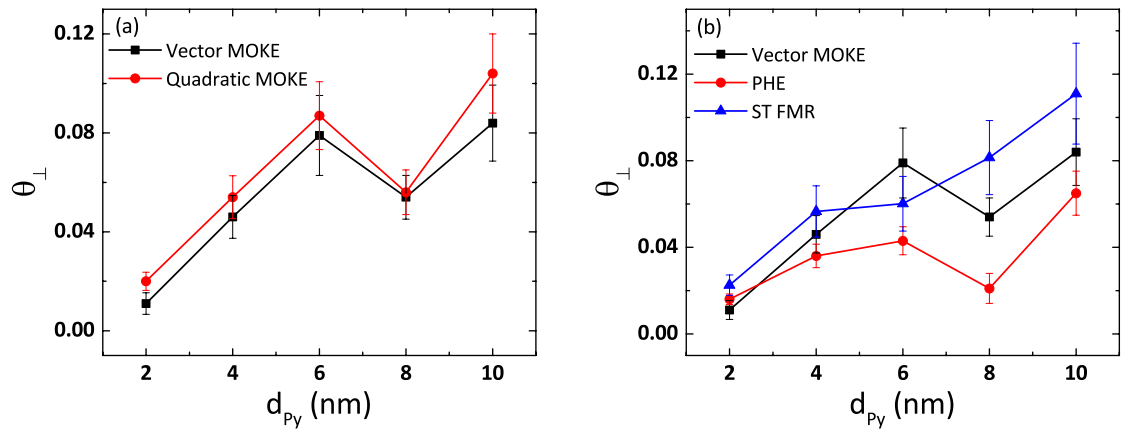

Figure 4. (a) Spin Hall angle measured with vector (black squares) and quadratic (red circles) MOKE vs. permalloy thickness $d_{Py}$. (b) Spin Hall angle measured with vector MOKE (black squares), PHE (red circles), and ST-FMR (blue triangles) vs. $d_{Py}$.

In conclusion, we have demonstrated a convenient 3D MOKE technique that can simultaneously measure the current-induced antidamping-like and field-like torques using normally-incident light. We have also implemented a self-calibration method to accurately determine the effective field. We find excellent agreement between the results of this technique and quadratic MOKE, ST-FMR, and PHE measurements for a series of Pt/Py bilayers with different Py thicknesses. The technique can be easily extended to measure spin-orbit torques in systems with perpendicular magnetization, as well as in systems with arbitrary magnetization direction. We anticipate this technique will be useful for further studies of current-induced magnetization reorientation in a variety of materials.

**Financial support**

This work is supported by DOE under grant number **DE-SC0016380** and NSF under grant number DMR-1624976This work is supported by DOE under grant number **DE-SC0016380** and NSF under grant number DMR-1624976

# 3D MOKE spin-orbit torque magnetometer

Halise Celik[1], Harsha Kannan[1], Tao Wang[1], Alex R. Mellnik[2*], Xin Fan[3], Xinran Zhou[4], Daniel C. Ralph[2], Matthew F. Doty[4], Virginia O. Lorenz[5] and John Q. Xiao[1†]

[1]*Department of Physics and Astronomy, University of Delaware, Newark, Delaware 19716, USA*

[2]*Department of Physics and Kavli Institute at Cornell for Nanoscale Science, Cornell University, Ithaca, New York 14853, USA*

[3] *Department of Physics and Astronomy, University of Denver, Denver, CO 80210, USA*

[4]*Department of Materials Science and Engineering, University of Delaware, Newark, Delaware 19716, USA*

[5] *Department of Physics, University of Illinois at Urbana-Champaign, Urbana, Illinois 61801, USA*

*Current Affiliation: Intel Corporation, Hillsboro, Oregon 97124, USA*

[†]jqx@udel.edu

## Supplementary Material

The magnitude of the SOT is determined through a self-calibration method. A line scan is performed by keeping the laser position fixed and translating the sample along the *y* direction as described in [1]. The difference between lock-in voltages at positive saturation field and negative saturation field are taken for the SOT signal and summation of the lock-in voltages at positive saturation field and negative saturation field are taken for the out-of-plane Oersted field for each position. Figure 3 shows a comparison of line scans obtained using a balanced detector for Py(4)/Pt(6) from Ref [2], and line scans from vector MOKE using quadrant detectors, indicating good agreement.

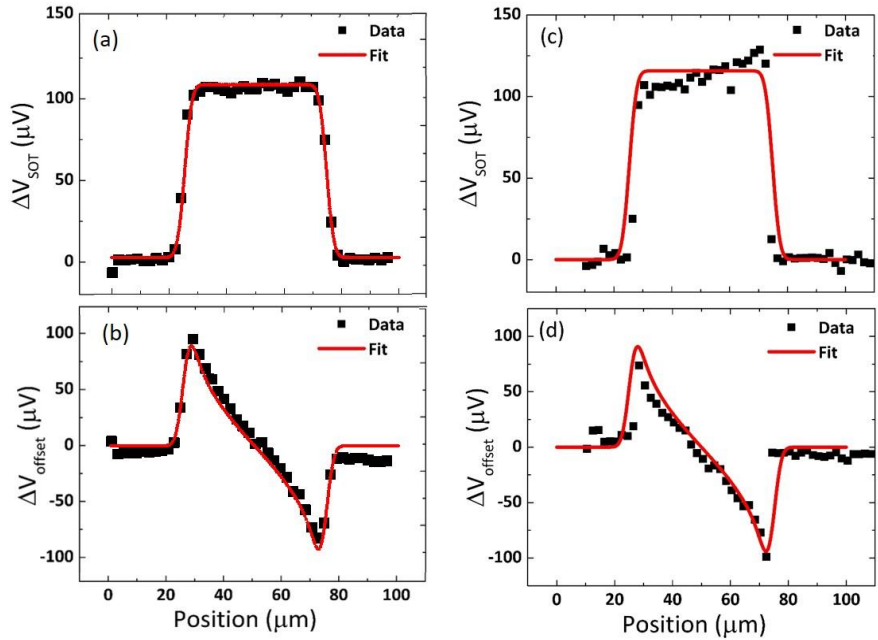

Figure 1. (a) Line scan result for Py(4)/Pt(6) with balanced detector. Out-of-plane equivalent spin-orbit field detected by subtracting signals taken at positive and negative saturation

field. Fit function (red line) is calculated as the integration of the SOT-induced magnetization reorientation weighted by the Gaussian function that describes the spatial distribution of the laser. (b) Out-of-plane Oersted field detected by addition of signals taken at positive and negative saturation field. The fit function for the Oersted field (red line) is similarly calculated as the integration of the local magnetization reorientation weighted by the Gaussian function that describes the spatial distribution of the laser. (c), (d) Line scan results for Py(4)/Pt(6) with quadrant detectors.